\documentclass[epj]{svjour}

\usepackage{graphicx}
\usepackage{fancyhdr}
\def\makeheadbox{{
 }}
\newcommand{\be}{\begin{eqnarray}}
\newcommand{\ben}{\begin{eqnarray}\nonumber}
\newcommand{\ee}{\end{eqnarray}}
\newcommand{\nee}{\nonumber \end{eqnarray}}

\def\mytitle{Transition to Exact Susy} 
\def\myauthors{L. Clavelli}    
\def\mytype{Contributed Talk}    
\def\mysession{Theoretical Models}

\begin{document}
\title{Extended Higgs Models and a Transition to Exact Susy}
\author{L. Clavelli
\thanks{\emph{Email:}lclavell@bama.ua.edu}%
}                     
\institute{
Department of Physics and Astronomy\\
University of Alabama\\
Tuscaloosa AL 35487\\ 
}
%
\date{}
\abstract{
   String landscape ideas and the observation of a positive vacuum energy
in the current universe suggest that there could be a future transition
to an exactly supersymmetric world.  Atomic and molecular binding in this
susy background probably require that electroweak symmetry breaking survives
the transition.  Among several susy higgs models that have been
discussed, one stands out in this regard.  Thus, the higgs structure that
is revealed at the LHC could have strong consequences for the type of
bulk matter that may arise in a future susy universe.  
} 
\maketitle
%

  The observations of a small but positive vacuum energy in our universe
plus the strong indications that, in its early moments, the universe made
transitions from states of much higher vacuum energy, raise the question
of whether there are further transitions to be expected within our 
horizon.  If so, it is natural to ask what properties this future universe
might have.
  
    We seem to be living in a bubble that formed some $13.7$ billion years
ago and that, after passing through many meta-stable states in a brief
inflationary era, transitioned to our current calm umiverse which is,
nonetheless, still inflating with a vacuum energy density measured to be 
\be
         \epsilon_{now} = 3.560 GeV/m^3 = (.0023 eV)^4 \quad .
\label{vacenergy}
\ee
This is some 124 orders of magnitude less than 
the natural value, $M_{Planck}^4$  that might have been expected 
for this quantity but it is known \cite{Weinberg} that arriving in
such a calm universe was a prerequisite for the evolution of advanced
life forms.  From a physics point of view it is, however, necessary
to ask what circumstances might have made this early 
history of our universe not extremely improbable.  
This seems to lead inevitably to speculation about 
possible regions of the universe outside of causal contact with us.
   
     For example, the scenario of
eternal inflation \cite{Linde} proposes that the universe is infinite
in spatial and temporal extent and that, consequently, however low
the probability of life is per unit of space-time volume, there are 
infinite numbers of civilizations in space-time that are similar to ours.  
This picture requires that there is an equilibrium established in the
``multiverse" and that the probability of ``jumping up" to a state of higher
vacuum energy is in statistical balance with that of ``jumping down".
It is also thought by many in this school that there are infinitely many
possible states of the universe that are massively anti-DeSitter, i.e.
possessing enormously negative vacuum energy density $\epsilon$.  
Such states would 
collapse in a big crunch on a time scale of $1/\sqrt{24 \pi G_N |\epsilon|}$.
The probability of a transition to such a deeply negative vacuum energy
should also be quite high raising the question of why our current universe
has persisted for its multi-billion year lifetime.     
Furthermore, naive physical intuition suggests that transitioning to
a state of lower energy density should be vastly more probable than
jumping up and it would seem that additional theoretical 
analysis would need to be done in this picture to establish a result 
preventing the 
universe everywhere from evolving inevitably to the lowest possible 
vacuum energy.  In any case, in the absence of experimental confirmation,
one is free to ask whether other possibilities exist.

     We study an alternate scenario in which the universe has a
supersymmetric (susy) gound state of exactly zero vacuum energy.  Examples of
such universes are provided by the five original superstring theories but
we prefer to think in terms of a simple supersymmetric extension of
the standard model. 
In this model the visible universe should eventually make a transition to the
susy ground state.  One could envision an inhomogeneous universe where,
outside of our horizon, shells of higher vacuum energy from the inflationary
era are inflating rapidly but without sufficient matter density to spawn
galaxies.  Note that in the standard false vacuum decay
theory \cite{Coleman}, vacuum energy goes into the bubble wall and not into creating
matter.  Outgoing shells may be unlikely to collide sufficiently to
create significant matter.

     Some attention has been given to the possible properties of a future
susy universe \cite{future,CL}.  The primary feature of such a universe is
a weakening of the Pauli Principle due to the degeneracy of fermions and
bosons.  Susy atoms, if they exist, would have entirely s-wave ground states.
As in our universe, quantum mechanics predicts that all binding energies
are proportional to the electron (or common electron/selectron) mass.
The mean radii of susy atoms would be inversely proportional to this mass.
Thus, unless electroweak symmetry breaking (EWSB) survives the transition
to the future susy universe providing masses, no electromagnetic bound 
states could be expected.

The time scale \cite{future} for the transition to take place is governed 
by the behavior of the cube of the cosmological scale factor
\be
      \tau = \sqrt{\frac{1}{24 \pi G_N \epsilon}} = 5.61 \cdot 10^9 yrs  \quad .
\ee

     We have investigated several popular phenomenological susy models 
from this point of view \cite{xhiggs}.  One could start from the most
general renormalizable superpotential, $W$, involving the mimimal pair of
higgs doublets plus a singlet field, $S$.  To avoid the 
uncontrolled growth of
MSSM related acronyms We refer to this most general
model with a single extra singlet as the 
Singlet Extended Susy Higgs Model (SESHM).
   
\be
    W = \lambda \left( S (H_u\cdot H_d - v^2) + \frac{\lambda^\prime}{3} S^3
           + \frac{\mu_0}{2}S^2 \right) \quad .
\ee

This leads to a scalar higgs potential with $F$ terms of the form
\ben
   V_F = \lambda^2 \left( |H_u\cdot H_d - v^2 + \lambda^\prime S^2 \right. \\
 \left. + \mu_0 S|^2   + |S|^2 (|H_u|^2 + |H_d|^2)\right) 
\label{VF}
\ee
and to $D$ terms of the form
\ben
     V_D &=& \frac{g_1^2+g_2^2}{8} \left( |H_d|^2 - |H_u|^2 \right)^2\\
&+& \frac{g_2^2}{2}\left( |H_d|^2 |H_u|^2 - |H_u \cdot H_d|^2\right)
\ee
where $g_1$ and $g_2$ are the $U(1)$ and $SU(2)$ gauge couplings.

There could also be susy breaking ``soft" higgs mass terms and other soft terms
proportional to terms in the superpotential.  These soft terms are something of
an embarrassment for susy theory since their origin is not understood.  The 
problem is often swept under the rug by referring to some unknown susy breaking
mechanism
in a ``hidden sector" which is communicated gravitationally or otherwise to our
sector.  However, interactions in string theory hidden sectors are similar to those
in our sector and, if we understood their origin in a hidden sector, the same 
mechanism should be available in our sector.  No treatment of susy breaking 
which appeals to soft susy breaking terms can be complete unless the origin of
these terms is also understood.

We would prefer to assume, as we
tentatively do here, that these
terms are absent unless forced on us for phenomenolgical reasons.  For example, in
the Minimal Susy Standard Model (MSSM) they are necessary to obtain consistency with
experimental constraints.  In any case, the soft terms will disappear if susy breaking
vanishes in both sectors though they are relevant in the broken susy phase.

The SESHM was first written some thirty years ago \cite{Fayet}.  
It was noted then that, unless the soft terms were somehow essential, the
susy breaking phase was at best meta-stable with the true minimum being exactly
supersymmetric.  Since then various sub-spaces of the $\lambda^\prime,\,\mu_0$
parameter space have been investigated under various names; see for example
\cite{Barger,reviews}.  

Some of these limiting cases and the corresponding labels are\\
 $v \,,\,\mu_0 \rightarrow 0 : \quad W \rightarrow NMSSM$, The ``Next to Minimal Susy Standard Model", \\
 $\lambda^\prime \,,\, \mu_0  \rightarrow 0: \quad W \rightarrow nMSSM$, the ``nearly Minimal Susy Standard Model"\\
 $\lambda^\prime \,,\, \mu_0 \,,\, v \rightarrow 0: \quad W \rightarrow UMSSM$, a $U(1)$ extended MSSM\\

It seems that the region of non-zero $\mu_0$ has not as yet been subjected to phenomenological
scrutiny, perhaps because this parameter can be consistently, though with some loss of generality, set to zero. 
Because of the symmetry between $H_u$ and $H_d$, the $D$ terms vanish at the minima.  At our present level
of investigation we have ignored phases, thus postponing the possibility of addressing CP violation.

The extrema of the potential define vacuum expectation values $v_0 =<H_u>=<H_d>$ and $S_0=<S>$ satisfying
\ben
     & \left.\frac{1}{\lambda^2}\frac{\partial V_F}{\partial S}\right|_{_0}=0\\
  =& (2 \lambda^\prime S_0 + \mu_0) (v_0^2-v^2 + \mu_0 S_0 + \lambda^\prime S_0^2) + 2 v_0^2 S_0
\label{conda}
\ee
\ben
     \left.\frac{1}{\lambda^2} \frac{\partial V_F}{\partial H_u}\right|_{_0} &=&0\\
 &=& v_0  (v_0^2-v^2 + \mu_0 S_0 + (\lambda^\prime +1) S_0^2 ) \quad .
\label{condb}
\ee

The solutions are\\
\noindent
{\bf Solution 1: $v_0 = v \quad , \quad S_0 = 0$}\\
This is one of the two grounds state of the model and corresponds to an exact supersymmetry
( vanishing vacuum energy) with EWSB ($v_0 \ne 0$).\\ 
\noindent
{\bf Solution 2: $v_0 = 0 \quad , \quad S_0 = \frac{-\mu_0 \pm \sqrt{\mu_0^2+
4 \lambda^\prime v^2}}{2 \lambda^\prime}$}\\
This solution is also supersymmetric with a vanishing vacuum energy at the
minimum but with no EWSB.\\
\noindent
{\bf Solution 3: $v_0 = 0 \quad , \quad S_0 = \frac{-\mu_0}{2 \lambda^\prime}$}\\
This corresponds to a broken susy with no EWSB.\\  
\noindent
{\bf Solution 4:} A solution to the minimization equations with non-zero $v_0$ and
$S_0$.
This corresponds to the phase most close to our current universe with 
broken susy and EWSB.  By eliminating $\mu_0 S_0$ from eqs.\,\ref{conda} and \ref{condb}
we find the conditions
\be
       S_0^2 = \frac{3 v_0^2 - v^2}{\lambda^\prime -1}
\label{S0sol}
\ee
and
\be
      \mu_0^2 = \frac{4 (\lambda^\prime v^2 - v_0^2 (2 \lambda^\prime + 1))^2}{(\lambda^\prime -1)(3 v_0^2 - v^2)} \quad .
\label{mu0sq}
\ee
These two equations determine $v_0$ and $S_0$ in terms of $\lambda^\prime$ and $\mu_0^2$.
One can see that eqs.\,\ref{S0sol} and \ref{mu0sq} each predict $v_0^2 > v^2/3$ for $\lambda^\prime>1$ and
$v_0^2 < v^2/3$ for $\lambda^\prime<1$.

For these extrema to be true minima, one would have to require that the parameters are such that the
eigenvalues of the matrix of the second derivatives of the potential are all positive at the
solutions of eqs.\ref{conda} and \ref{condb}.  
Since the potential is positive definite or zero, any localized solution
with vanishing vacuum energy, $V_F(0)$, is necessarily a minimum.  
For the broken susy case, we must study
the mass squared matrix
\ben
   M^2 = \left| \begin{array}{ccc}    \quad  \alpha+\zeta \quad & \quad \gamma-\zeta \quad & \quad \delta \quad\\
        \quad  \gamma-\zeta  \quad & \quad \alpha + \zeta \quad & \quad \delta \quad\\
        \quad \delta  \quad      &  \quad \delta  \quad       & \quad \beta \quad \end{array} \right|
\ee
where
\be
     \alpha = \left.\frac{\partial^2 V_F}{\partial H_u^2}\right|_{0} = 2 \lambda^2 (v_0^2 + S_0^2)
\ee
\be
     \gamma = \left.\frac{\partial^2 V_F}{\partial H_u \partial H_d}\right|_{0} = 2 \lambda^2 (v_0^2 - S_0^2)
\ee
\be 
     \delta = \left.\frac{\partial^2 V_F}{\partial H_u \partial S}\right|_{0} = 2 \lambda^2 v_0 \left(2(
 \lambda^\prime + 1) S_0 + \mu_0 \right)
\ee
\ben
     \beta = \left.\frac{\partial^2 V_F}{\partial S^2}\right|_{0} &=& 2 \lambda^2 \left(-2 \lambda^\prime S_0^2\right.\\
            &+& \left.(2 \lambda^\prime S_0 + \mu_0)^2 + 2 v_0^2\right) 
\label{beta}
\ee
\be
     \zeta =  \left.\frac{\partial^2 V_D}{\partial H_u^2}\right|_{0} = (g_1^2+g_2^2) v_0^2 \quad .
\ee 
The $g_i$ are the electroweak coupling constants and $\zeta$ is proportional to the $Z$ mass squared.   
The eigenvalues of the mass squared matrix satisfy
\be
     m_1^2 = 4 \lambda^2 S_0^2 + 2 \zeta \quad ,
\ee
\be
     m_2^2 + m_3^2 = \alpha + \beta + \gamma \quad ,
\ee
and 
\be
     m_2^2 m_3^2 = \beta (\alpha + \gamma) - 2 \delta^2 \quad .
\ee
The first squared mass is positive definite and corresponds to the eigenvector
\be
     \Psi_1 = \frac{H_u-H_d}{\sqrt 2} \quad .
\ee

The positivity of $m_2^2$ and $m_3^2$ puts constraints on the parameter space of
$\lambda^\prime$ and $\mu_0$,  namely
\be
    m_2^2+m_3^2 = 4 \lambda^2  v_0^2 + \beta > 0
\label{sum}
\ee
and
\ben
    m_2^2 m_3^2 &=& \frac{16 \lambda^4 v_0^2}{\lambda^\prime-1}\left(-v_0^2 (6 \lambda^\prime + 3)+
v^2 (\lambda^\prime +2)\right)\\
     &>& 0 \quad .
\label{prod}
\ee
A simultaneous solution of eqs.\,\ref{sum} and \ref{prod} requires that $\lambda^\prime < -2$
and that 
\be
v_0^2 < v^2 \frac{\lambda^\prime + 2}{6 \lambda^\prime + 3} \quad .
\ee
For example, it is clear from eq.\,\ref{beta} that negative $\lambda^\prime$ guarantees the
positivity of eq.\,\ref{sum}.  It requires some analysis to see that no solutions with
positive $\lambda^\prime$ exist except, perhaps, over a set of measure zero.
Thus the broken susy phase with EWSB lies between the two exact susy phases.  Assuming
equality of the Yukawa couplings in the various phases, the common electron and selectron mass
in the exact susy phase with EWSB would be $v/v_0$ times greater than in the broken susy
phase.  Thus the transition would be endothermic requiring an extra input of energy.
For elements above helium, this energy could come from the energy released from the Pauli
towers as nucleons convert to scalar nucleons.

An interesting result, emphasized in \cite{xhiggs}, is that the NMSSM and the UMSSM, like the MSSM, 
have no EWSB in the susy limit
(i.e. $v_0 =0$ in the susy phase).  Only models with non-zero $v$ offer possibilities for atomic 
and molecular structures in a future susy phase.  In addition, only models with non-zero $\mu_0$
allow a true susy breaking minimum in the absence of soft masses. 

\begin{figure}
\includegraphics[width=0.45\textwidth,height=0.4\textwidth,angle=0]{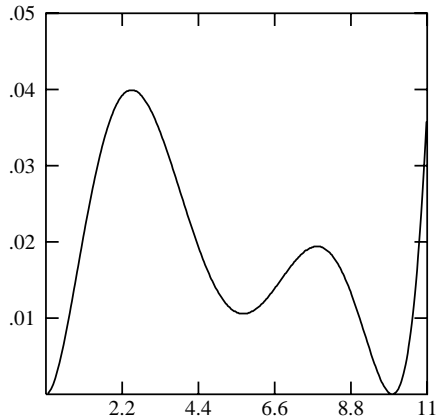}
\caption{The effective potential as a function of the magnitude of the
doublet higgs field showing qualitatively the false vacuum of broken susy with EWSB and the
two degenerate vacua of exact susy.  Numerical values are given in arbitrary units.}
\label{potential}
\end{figure}

At our current level of analysis
neglecting possible yukawa coupling changes, soft susy breaking terms and phases in the
potential,  the higgs structure is as in figure\,\ref{potential} so
there is no exothermic transition to the exact susy minimum with EWSB. 

If higgs studies at the Large Hadron Collider (LHC) confirm
an extended higgs structure with non-zero $v$,  one could be encouraged to imagine 
the rise of an exactly supersymmetric civilization in the distant future
providing the Yukawa couplings decrease in the transition to exact susy or some
extra source of energy input is found.  
On the other hand, if the
LHC confirms the MSSM or an extended higgs model with $v=0$ One might
expect a structureless future susy universe with no atomic or molecular physics.

    The mass squared matrix discussed above allows positive higgs masses even in the absence of
soft breaking terms.  Current limits on higgs masses will further constrain the viable parameter space
although, because of the new interactions with the singlet higgs, one must avoid blindly applying the
published limits on higgs masses.  
This is contrary to the case of the MSSM where soft higgs masses are essential.  
If, however, future
experimental constraints require soft higgs masses even in the SESHM, there could be
a parameter space region allowing
an exothermic transition to an exact susy universe with EWSB.  In principle, 
with sufficient experimentation at CERN, the effects of the soft breaking masses can be disentangled
and the question of whether or not there could be an exothermic transition to the 
exact susy state with EWSB could be re-analysed.
Another broken susy minimum with no EWSB might also exist corresponding to solution three discussed above.

     The second susy ground state (solution 2), appearing at the origin in figure \ref{potential} 
has no EWSB.  In this ground state, matter exists only in a permanently ionized state with no possibility of
electrons condensing onto nuclei.  

    A question raised at this conference is whether this work is related to the anthropic principle (AP).  
The answer is no or, at least, not yet.  Physical principles summarize a large body of experimental 
observation.  A good example is Hamilton's principle of least action.  The AP is the observation that,
within narrow bounds, the laws of physics are such as to allow the evolution of intelligent life.
Nothing in the large body of current experimental observations leading to the AP suggests that 
intelligent life will also arise in a future phase of the universe such as the exact susy phase
discussed here.  The anthropic principle cannot be used to predict the
future. The fact that we exist does not prove that life forms of comparable (limited) intelligence 
will exist in the distant future.  In the work reviewed here, there is no prediction as to whether or not
the future susy phase will be friendly to advanced life forms.  Instead we merely point out a
possible correlation between parameters of the higgs potential measurable in the near future to
possibilities for life or their absence in a distant future susy universe.

    On the other hand, if intelligent life forms do evolve in some future atomphilic susy phase, those
beings might be able to deduce that their existence implied a prior broken susy phase.  For example,   
if the universe had been born in a susy phase of zero vacuum energy there might not have been sufficient
structure formation in the very early universe or the universe might have collapsed in the early instants.
Also, we know that CP violation was crucial in the production of the baryon asymmetry which, in turn,
seems essential to the development of stars, planets, and advanced life forms.  
There seems to be insufficent CP violation in the standard model so it is widely expected that 
susy CP violation played an essential role.  Susy CP violation, however, is presently thought to come 
from the non-zero
relative phases of the soft breaking terms and the $\mu$ parameter such as might exist in a broken
susy phase but not in an exact susy phase.  In the SESHM, as discussed
above, there might be no soft breaking terms but CP violation in the broken susy phase 
could come from a non-zero relative phase of the
vevs of $S$ and $H$.  These also would disappear in the exact susy phases (solutions 1 and 2 above)
where $S_0 v_0 = 0$.

    Subsequent to the work described here \cite{xhiggs}, a paper \cite{Profumo} has appeared applying 
similar inter-phase arguments to the EWSB phase transition of the early universe.
There is at present a rapidly growing body of work studying other aspects of metastable susy breaking.

{\bf Acknowledgements}
    This work was supported in part by the US Department of Energy under
grant DE-FG02-96ER-40967.   We are grateful for the
hospitality of the Institute for Fundamental Theory at the University
of Florida during the winter quarter of 2007 when most of the results
of this paper were extracted.  Some observations of Salah Nasri at
the University of Florida were key to the directions taken in this work.  
We also acknowledge the hospitality of the INFN at the
University of Bologna and several discussions on the topics of this paper
with the particle theory group there in the spring of 2007.  


\begin{thebibliography}{999}
\bibitem{Weinberg} S. Weinberg, Phys. Rev. Lett. {\textbf 59}, (1987) 2607
\bibitem{Linde} A. Linde, ArXiv:0705.1160
\bibitem{Coleman} S. Coleman, { Phys. Rev. D} {\textbf 15}, (1977) 2929 ;C.G. Callan and S. Coleman, { Phys. Rev. D} {\textbf 16}, (1977) 1762 ; S. Coleman and Frank DeLuccia, { Phys. Rev.D} {\textbf 21}, (1980) 3305  
\bibitem{future} L. Clavelli, Int. J. of Mod. Phys. E {\textbf Vol 15}, No. 6, (2006) 1-17
\bibitem{CL} L. Clavelli and  T. Lovorn  Int. J. of Mod. Phys. {\textbf A22} No. 12, (2007) 2133-2144
\bibitem{xhiggs} L. Clavelli, ArXiv:0705.1290
\bibitem{Fayet} P. Fayet, {Nucl. Phys.} B{\textbf 113}, (1976) 135;
{Phys. Lett.}B{\textbf 78}, (1978) 417
\bibitem{Barger} V. Barger, P. Langacker, H.S. Lee, and G. Shaughnessy,
{Phys. Rev. D} {\textbf 73}, (2006) 115010 
\bibitem{reviews} S. Kraml et al., hep-ph/0608079\\
V. Barger, P. Langacker, and G. Shaughnessy, hep-ph/0611112,
Proceedings of the 14th Int. Conf. on Supersymmetry and the Fundamental
Interactions, Irvine CA, 12-17 June, 2006, J. Feng ed., AIP 2007
\bibitem{Profumo} S. Profumo, M.J. Ramsey-Musolf, and G. Shaughnessy,
ArXiv:0705.2425
\end{thebibliography}
\end{document}